\documentclass[AMA,STIX1COL]{WileyNJD-v2}
\usepackage{moreverb}
\usepackage{subcaption}
\usepackage{color,soul}

\newcommand\BibTeX{{\rmfamily B\kern-.05em \textsc{i\kern-.025em b}\kern-.08em
T\kern-.1667em\lower.7ex\hbox{E}\kern-.125emX}}

\articletype{Article Type}%

\received{<day> <Month>, <year>}
\revised{<day> <Month>, <year>}
\accepted{<day> <Month>, <year>}

\begin{document}

\title{Introducing Hann windows for reducing edge-effects in patch-based image segmentation}

\author[1]{Nicolas Pielawski} 

\author[1,2]{Carolina W\"ahlby} 

\authormark{Pielawski \textsc{et al}}

\address[1]{\orgdiv{Dept. of Information Technology}, \orgname{Uppsala University}, \orgaddress{\state{Uppsala}, \country{Sweden}}}

\address[2]{\orgdiv{BioImage Informatics Facility}, \orgname{SciLifeLab}, \orgaddress{\state{Uppsala}, \country{Sweden}}}

\corres{*Nicolas Pielawski, Department of Information Technology, Polacksbacken (L\"{a}gerhyddsv\"{a}gen 2), Uppsala. \email{nicolas.pielawski@it.uu.se}}

\presentaddress{Department of Information Technology, Polacksbacken (L\"{a}gerhyddsv\"{a}gen 2), Uppsala}

\abstract[Abstract]{There is a limitation in the size of an image that can be processed using computationally demanding methods such as e.g. Convolutional Neural Networks (CNNs). Some imaging modalities -- notably biological and medical -- can result in images up to a few gigapixels in size, meaning that they have to be divided into smaller parts, or patches, for processing. However, when performing image segmentation, this may lead to undesirable artefacts, such as edge effects in the final re-combined image. We introduce windowing methods from signal processing to effectively reduce such edge effects. With the assumption that the central part of an image patch often holds richer contextual information than its sides and corners, we reconstruct the prediction by overlapping patches that are being weighted depending on 2-dimensional windows. We compare the results of four different windows: Hann, Bartlett-Hann, Triangular and a recently proposed window by Cui et al., and show that the cosine-based Hann window achieves the best improvement as measured by the Structural Similarity Index (SSIM). The proposed windowing method can be used together with any CNN model for segmentation without any modification and significantly improves network predictions.}


\jnlcitation{\cname{%
\author{N. Pielawski}, and
\author{C. W\"ahlby}} (\cyear{2019}), 
\ctitle{Introducing Hann windows for reducing edge-effects in patch-based image segmentation}, \cjournal{}, \cvol{}.}

\maketitle


\section{Introduction}\label{sec1}

Semantic image segmentation is a process consisting of separating an image into regions, e.g. representing different object types. This problem is ill-defined because there is no general definition of a region, and learning-based methods such as CNNs have started to outperform classical rule-based methods in recent years. In 2015 \citet{ronneberger2015u} introduced the U-Net neural network and architecture, consisting of a compressing and decompressing part with skip-connections in between. In 2017, \citet{jegou2017one} greatly increased the number of skip-connections used, and reached the state-of-the-art of semantic segmentation with very few trainable parameters, but consequently at the cost of a larger memory footprint. 

Segmentation tasks are memory intensive, mainly due to the size of the input images and the preservation of feature maps along the computational graph of a CNN. In some cases, such as when handling giga-pixel-sized whole slide images (WSI) in digital pathology, it is not possible to process the whole image at once\cite{solorzano2018whole}. To counteract these memory issues, patch-based segmentation methods use different techniques that feed more or less contextual information to the neural networks. Edge-effects are all known to appear when working with CNNs, and have been approached in different ways: for instance by keeping a contextual border in the input that is removed from the output in the original U-Net architecture. In 2018, \citet{innamorati2018learning} showed that the errors of segmentation are higher for the pixels near the edges and even worse for the corners. The same year, \citet{cui2018deep} proposed a method to reduce edge effects and increase the final segmentation quality after reassembling the different patches. Their method consists of weighing the loss function and the patches with a specific mask which will be referred to as Pyramidal window in this paper.

In signal processing, one of the renowned windows is the Hann window, invented by Julius von Hann around 1900 and named after him by Blackman and Tukey\cite{blackman1958measurement} in 1958. Window functions are often used to taper a signal, by multiplying a window by a patch extracted from the signal, reducing the importance of the borders. Their usage is broad, for instance when transforming a signal into a spectrogram that can later reconstruct the original signal without artefacts. Another example lies in the field of statistics where curve fitting can be given a weighing factor --- the window function --- which is also known under the name of kernel. It is desirable for these windows to integrate to one (with a stride of half a window); not filling the criterion requires an additional normalisation step after reconstruction which is the case of the Pyramidal window. This supplementary step can introduce additional artefacts due to rounding errors in floating point arithmetic.

In this paper we explore the idea of windowing to reduce edge effects after CNN-based image segmentation. Our contributions are the following:
\begin{itemize}
    \item We present a method that can be applied \textit{post hoc} on any CNN segmentation output without needing to retrain or modify the loss function nor renormalise the output.
    \item We handle edge and corner cases separately, which gives a weighted estimate given the available context. We also avoid additional floating point errors by using windows that integrate to one.
\end{itemize}

\section{Method}

The proposed window patch-based method is a refinement step that reduces the edge artefacts at patch borders. Inspired by signal processing, we multiply each patch with a 2-dimensional window function, which gives more emphasis to their centres and less to their adjacent edges and corners. Next, we hypothesise that the most correct information will be kept when combining the window-weighted CNN outputs increasing the quality of our predictions.

Our method follows this pipeline:
\begin{enumerate}
    \item Extracting overlapping patches, with a stride of half a patch size.
    \item Performing the prediction on the patches.
    \item Multiplying each patch by the appropriate window depending on its absolute location: the window must be of the same size of the patch and must be replaced if it is associated to a border or corner patch.
    \item Summing all the patches at their absolute location.
\end{enumerate}

\subsection{Window functions}
We evaluate three different windows from classical signal processing: Hann\cite{ha1989new}, Bartlett-Hann\cite{ha1989new}, and Triangular\cite{harris1978use} and compare with the Pyramidal window\cite{cui2018deep}. We chose to focus on evaluating these windows rendered in 2 dimensions. The choice of a window function is arbitrary for as long as it is separable and sums up to 1 when properly integrated over the patches with a stride of half a window size. The Pyramidal window does not follow this requirement and thus needs an extra normalisation step which can introduce additional artefacts.

In signal processing, for an arbitrary 1-dimensional window function $w$, \citet{speake1981note} described the 2-dimensional version $W$ in separable form as:
$$W(i,j)=w(i)w(j)$$

We can thus derive the 2-dimensional versions of the original windows as follows:

\begin{equation}
    W_{Average}(i,j)={\frac{1}{4}}\;
\end{equation}

\begin{equation}
    W_{Hann}(i,j)={\frac{1}{4}}\;\left(1-\cos \left({\frac {2\pi i}{I-1}}\right)\right)\left(1-\cos \left({\frac {2\pi j}{J-1}}\right)\right)
\end{equation}

\begin{equation}
    W_{Bartlett-Hann}(i,j)=\left(a_0+a_1\left|\frac{i}{I}-\frac{1}{2}\right|-a_2\cos(\frac{2\pi i}{I}\right)\left(a_0+a_1\left|\frac{j}{J}-\frac{1}{2}\right|-a_2\cos(\frac{2\pi j}{J}\right)
\end{equation}

\begin{equation}
    W_{Triangular}(i,j)=\left(1-\left|\frac{2i}{I}-1\right|\right)\left(1-\left|\frac{2j}{J}-1\right|\right)
\end{equation}

with $i$ the current horizontal position, $I$ the width of the patch, $j$ the current vertical position, $J$ the height of the patch, $a_0 = 0.62$, $a_1 = 0.48$, and $a_2 = 0.38$\cite{ha1989new}.

\citet{cui2018deep} defined a weighted loss function which is later used as a window. The resulting Pyramidal window is defined as follows:
\begin{align}
    W_{Pyramidal}(i,j)&=\alpha\frac{D_{i, j}^{e}}{D_{i, j}^{c} + D_{i, j}^{e}} \\
    \alpha&=\frac{I \cdot J}{\sum_{i=1}^{I}{\sum_{j=1}^{J}{\frac{D_{i, j}^{e}}{D_{i, j}^{c} + D_{i, j}^{e}}}}} \nonumber
\end{align}

with $D_{i, j}^{e}$ the absolute distance from the edge and $D_{i, j}^{c}$ the distance from the centre. The Pyramidal window formula above cannot be described in separable form which consequently increases the cost of computing the 2-dimensional window.

\begin{figure}[ht]
    \centering
    \includegraphics[width=\columnwidth]{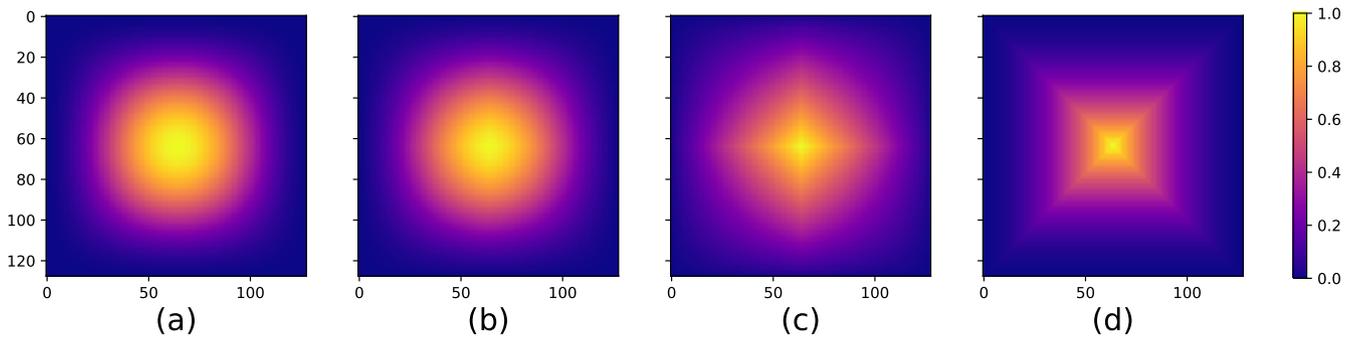}
    \caption{Illustration of the different 2-dimensional windows. Every window gives more emphasis to the information in the centre than the information on the borders and in corners. \textbf{(a)} 2D Hann window, \textbf{(b)} 2D Bartlett-Hann window, \textbf{(c)} 2D Triangular window, and \textbf{(d)} 2D Pyramidal window.}
    \label{fig1}
\end{figure}

The 2-dimensional realisation of the different windows can be seen in figure \ref{fig1}.

\subsubsection{Complexity}
The complexity of a non-overlapping reconstruction takes $nm$ computations of individual patches, with $n$ the number of patches horizontally, and $m$ the number of patches vertically. Our method has a complexity of $$\left(2n-1\right)\left(2m-1\right)=4nm-2n-2m+1$$
That is approximately $4nm$, and this approximation becomes more accurate as $n$ or $m$ grows, i.e., the image gets larger.

In practice, the non-overlapping reconstruction can be performed first in order to display a preview to an user, and then complete the missing information by computing the three quarters of the remaining patches in the meanwhile. Another optimisation is to combine both approaches and use the fast non-overlapping reconstruction method for the inessentials details of an image (e.g. the background) and the windowed method for the objects of interest.

\subsection{Edges and corners}
Contextual information from adjacent patches is naturally not available at the edges and corners of the full-size image. Therefore we propose specific windows to increase inference accuracy in these parts of the final image. The patches nearby the edges and corners are weighted with a different set of windows to compensate for the missing information so that the sum of all overlapping windows over the full image is 1.
Border patches of an image are defined with the following equations, and work with any of our proposed windows:

\begin{align}
W_{Up}(i,j)&= 
\begin{cases}
    w(i),& \text{if } j<\frac{J}{2} \\
    w(i)w(j),& \text{otherwise}
\end{cases}\\
W_{Down}(i,j)&= 
\begin{cases}
    w(i),& \text{if } j>\frac{J}{2} \\
    w(i)w(j),& \text{otherwise}
\end{cases}\\
W_{Left}(i,j)&= 
\begin{cases}
    w(j),& \text{if } i<\frac{I}{2} \\
    w(i)w(j),& \text{otherwise}
\end{cases}\\
W_{Right}(i,j)&= 
\begin{cases}
    w(j),& \text{if } i>
    \frac{I}{2} \\
    w(i)w(j),& \text{otherwise}
\end{cases}
\end{align}
with $w(i)$ the evaluation of an arbitrary window $w$ at position $i$.

The formula of the upper left corner patch of an image is defined as:

\begin{equation}\label{eq:corner}
    W_{UpLeft}(i,j)= 
\begin{cases}
    1,& \text{if } \leq\frac{I}{2}\;and\;j\leq\frac{J}{2} \\
    w(i),& \text{if } i>\frac{I}{2}\;and\;j<\frac{J}{2} \\
    w(j),& \text{if } i<\frac{I}{2}\;and\;j>\frac{J}{2}\\
    w(i)w(j),& \text{otherwise}
\end{cases}
\end{equation}

It is possible to construct the three remaining corner windows from \ref{eq:corner}.

\begin{figure}
    \centering
    \includegraphics[width=80mm]{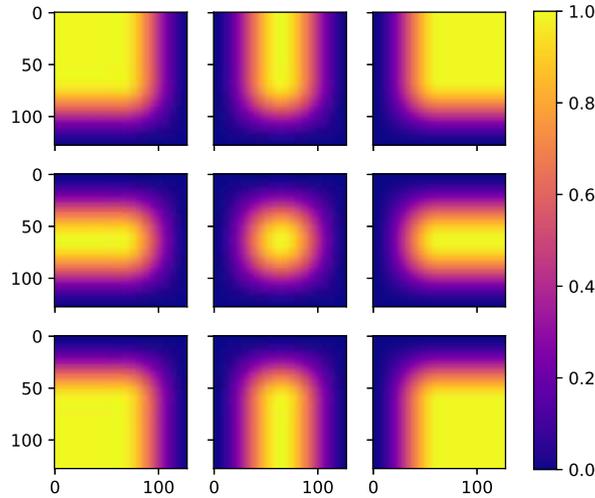}
    \caption{The different configurations of Hann windows of size 128x128 for edge and corner cases. Summing all the windows with the appropriate amount of overlap ($\frac{128}{2}=64$ pixels in this example) makes every window applied to the reconstructed image sum to 1.}
    \label{fig2}
\end{figure}

These formulas yield eight new windows as visualised in Figure \ref{fig2}.

\section{Experiments}
We present an experiment where five different approaches for combining the results of patch-based image segmentation are compared. The experiments follow the same protocol and data as in \citet{cui2018deep}, where a U-Net neural network architecture is trained on a hematoxylin and eosin (H\&E)-stained tissue dataset. The model is trained with a cross-entropy loss, and the weighting loss as constructed by \citet{cui2018deep} was discarded. The patch size is 128x128 for a complete image of 1024x1024 pixels. The five different approaches that are compared are: Average, Pyramidal, Hann, Bartlett-Hann, and Triangular windows.

\begin{figure}[ht]
    \centering
    \includegraphics[width=\columnwidth]{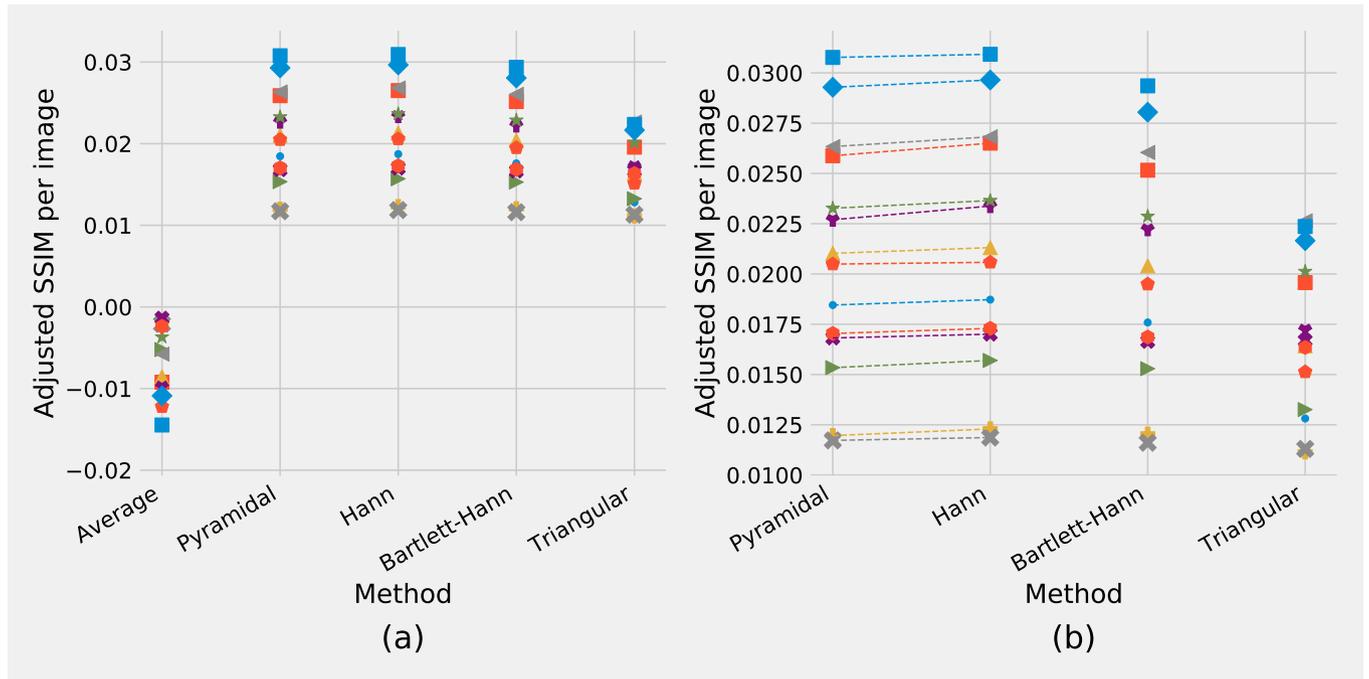}
    \caption{\textbf{(a)} and \textbf{(b)} show the adjusted SSIM for six different methods and a sample size of 14 images. Each image has a unique combination of symbol and colour. The SSIM indexes have been computed channel-wise then averaged. The baseline (the SSIM from non-overlapping patches) has been subtracted from each score. \textbf{(a)} shows the five different windows that we compared in this study. \textbf{(b)} shows the different methods without the average window in order to emphasise the differences. The differences between the Pyramidal and the Hann windows are not clearly visible and dashed lines have been plotted to display a visual trend.}
    \label{fig3}
\end{figure}

After reconstructing the output image we computed the scores using the structural similarity index\cite{wang2004image} (SSIM) using the image segmentation ground truth and the predictions. The prediction consisted of three classes, and we calculated the SSIM for each of the three predictions, and then averaged in order to achieve the resulting scores. The SSIM indexes have been computed channel-wise then averaged. Due to a low variance in the SSIM indices for each method but a high variance between methods, we decided to subtract the SSIM of the baseline, i.e. the non-overlapping patches with the most prominent artefacts, to the SSIM of each method. Figure \ref{fig3} shows the adjusted SSIM for each experiment and each image. Each marker is representing an image with a unique combination of symbol and colour.

\subsection{Hypothesis testing}

Besides the average window, all windows significantly outperformed the baseline (using adjacent patches, without overlaps). An independent paired t-test was conducted and showed evidence that the Pyramidal window (M=0.02079, SD=0.00575; t(13)=13.0294, p<.0001), Hann window (M=0.02112, SD=0.00581; t(13)=13.1095, p<.0001), Bartlett-Hann window (M=0.02026, SD=0.00542; t(13)=13.4807, p<.0001) and Triangular window (M=0.01691, SD=0.00379; t(13)=16.0797, p<.0001) all yielded a better score than the baseline.

Surprisingly, predictions with overlapping patches weighted by an average window (M=-0.00710, SD=0.00431) performed worse than our baseline; t(13)=5.9431, p<0.0001, even though they contain about four times as much information. This most likely happens because the overlapping patches do not weigh down edge artefacts thus yielding four times as many artefacts.

In our results, the Hann outperformed the Pyramidal window by a small amount and we performed an exact sign test\cite{dixon1946statistical} in order to highlight those differences in SSIM. The Hann window elicited a statistically significant mean increase in SSIM ($0.00033 \pm 0.00017$) compared to the Pyramidal window; p<.0002. 

\subsection{Visual results}
\begin{figure}[ht]
    \centering
    \begin{subfigure}{0.24\linewidth}
	    \includegraphics[width=42mm]{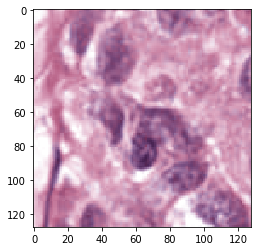}
    	\caption{}
    \end{subfigure}
    \begin{subfigure}{0.24\linewidth}
	    \includegraphics[width=42mm]{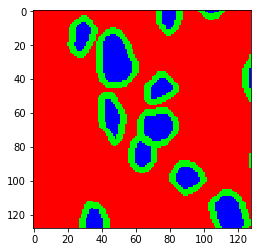}
    	\caption{}
    \end{subfigure}
	\begin{subfigure}{0.24\linewidth}
	    \includegraphics[width=42mm]{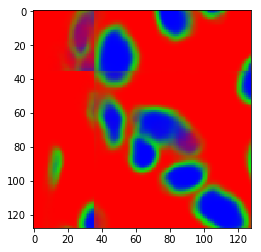}
	    \caption{}
    \end{subfigure}
	\begin{subfigure}{0.24\linewidth}
	    \includegraphics[width=42mm]{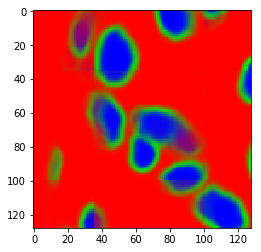}
	    \caption{}
    \end{subfigure}
	\begin{subfigure}{0.24\linewidth}
	    \includegraphics[width=42mm]{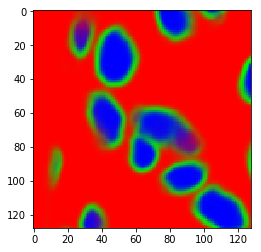}
	    \caption{}
    \end{subfigure}
	\begin{subfigure}{0.24\linewidth}
	    \includegraphics[width=42mm]{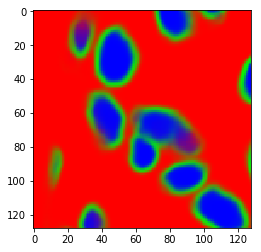}
	    \caption{}
    \end{subfigure}
	\begin{subfigure}{0.24\linewidth}
	    \includegraphics[width=42mm]{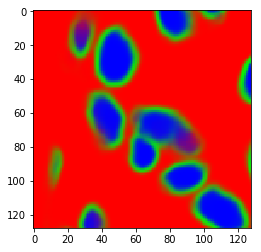}
	    \caption{}
    \end{subfigure}
	\begin{subfigure}{0.24\linewidth}
	    \includegraphics[width=42mm]{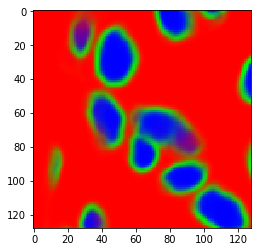}
	    \caption{}
    \end{subfigure}
    \caption{Input, ground-truth and different reconstruction methods. The figure is cropped from 1024x1024 to 128x128 for visibility. \textbf{(a)} The input grayscale image. \textbf{(b)} The ground-truth. \textbf{(c)} The baseline, where the patches are assembled without overlaps. \textbf{(d)} The overlapping patches assembled with an average window. \textbf{(e)} The patches assembled and weighted with a Pyramidal window. \textbf{(f)} The patches assembled and weighted with a Hann window. \textbf{(g)} The patches assembled and weighted with a Bartley-Hann window. \textbf{(h)} The patches assembled and weighted with a Triangular window.}
    \label{fig4}
\end{figure}

Figure \ref{fig4} shows the results of the different methods and can be compared to the input and ground truth. Although subtle, artefacts are most visible in the baseline and the average window versions, the visual differences in the remaining results remain marginal.

\captionsetup[subfigure]{font=scriptsize,labelfont=scriptsize}

\section{Discussion and future work}

Our method offers an immediate reduction in edges artefacts and can be easily implemented and integrated without the need of modifying any existing Deep Learning model. The theory stemming from signal theory provides a basis for further improvements. Our results suggest that using a Hann window is an effective way of reducing edge artefacts, and testing the method on different datasets and image modalities could further confirm our hypothesis. More window types could be compared and a more in-depth study could focus on determining the best window type depending on the circumstances.

The proposed method assumes that a constant amount of context is needed to have an accurate prediction. It could be of interest to focus on reducing the amount of context needed, which would result in saturating windows. For instance, the Tukey window\cite{harris1978use}, has a tunable parameter varying the amount of context. This would result in reducing the amount of overlap between the patches and could bring important computational gains.

An optimisation of the context could also be achieved with Deep Bayesian neural networks that yield a prediction associated to an uncertainty. This uncertainty, or precision, could be combined to a Bayesian prior in order to compute a different window for each patch. This window would then be predicted depending on how much context the neural network needs: the more certain a predicted area will be, the more weight it will receive.

\section{Conclusion}
In this paper we describe a new method that introduces Hann windows for reducing the edge-effects when performing image segmentation with CNNs. We explained how to construct arbitrary windows in 2-dimensions and how they can be expanded for borders and corner cases. To demonstrate our concept, we tested five different windows on a cell nuclei dataset and showed that it compares favourably with an existing method provided by \citet{cui2018deep}. Finally, the method is readily available and simple to implement in existing Deep Learning models, even if they are already trained.

\section{Acknowledgements}
This project was financially supported by the Swedish Foundation for Strategic Research (grant SB16-0046).

\end{document}